# Reimagining Internet Geographies:

# A User-Centric Ethnological Mapping of the World Wide Web

*Angela Xiao Wu & Harsh Taneja*[1]

**Suggested Citation:**

## Abstract

We propose a new user-centric imagery of the WWW that foregrounds local usage and its shaping forces, in contrast to existing imageries that prioritize Internet infrastructure. We construct ethnological maps of WWW usage through a network analysis of shared global traffic between 1000 most popular websites at three time points and develop granular measures for exploring global participation in online communication. Our results reveal the significant growth and thickening of online regional cultures associated with the global South. We draw attention to how local cultural identity, affirmative state intervention and economic contexts shape regional cultures on the global WWW.

[1] Both authors contributed equally to this work. James Webster and Wendy Griswold provided generous feedback on earlier drafts and Jan Servaes pointed to relevant literature. Acknowledgments are also due to Sushant Tripathy for his programming support and Manu Bhandari for proofreading the draft at a short notice. This project was made possible by the research grants from the Internet Policy Observatory at the University of Pennsylvania, the Journalism Library of the University of Missouri, and the C-Centre at the School of Journalism and Communication at the Chinese University of Hong Kong

## Reimagining Internet Geographies:

## A User-Centric Ethnological Mapping of the World Wide Web

Our study is a conceptual and methodological endeavor to bring about newer imageries of the World Wide Web (WWW) based on web usage. This approach signals a shift in perspective from prioritizing Internet infrastructure[1] (including technical connectivity) to focusing on local user engagements. We empirically account for local communication flows enabled by the WWW, making visible, on an unprecedented scale, aspects of Internet use that remain "largely invisible" when "viewed from the perspective of network centers" (Burrell, 2012).

To achieve this, we develop a research method that we call "user-centric ethnological mapping of the WWW," which empirically examines global web usage in terms of the scale, cultural distance and thickening of local communication landscapes or "online regional cultures." We implement this mapping by analyzing shared audience traffic between the world's top 1000 sites, which generally draw over 99% of total web traffic, in 2009, 2011 and 2013. Our analyses show significant changes in global web use. Specifically, changes in the composition of various online regional cultures as well as their size, distance and thickness scores—measures we develop and compute in this study—demonstrate the rise of the global South along the decentralization or de-Americanization of the WWW. Our study challenges the predominant imaginations of the WWW and contributes to a fuller understanding of actual user behavior online by documenting the recent development of the WWW into distinct but interconnected regional cultures. We highlight the possible role of local cultural identity and political economy, particularly specific types of government interventions, in this evolution.

## Reimagining the WWW: From Technical Features to User Practices

Searching for the images of the "Internet" on the Internet, one sees myriad portrayals, some artistic illustrations and some based on certain empirical evidence. These maps reveal "architectural plans, engineering blueprints, anatomical drawings, and statistical graphics;" all attempt at providing a factual reality of the Internet (Dodge, 2008, p 352). Apart from providing fascinating diverse images of the Internet, they provide insights into "how people imagine the virtual territory" that aligns with certain "interests and desires" (Dodge, 2008, p. 13). More importantly, popular representations of the Internet, like representations of other kinds of space, may function to legitimize the status quo without our being aware of it (Graham, 2014).

Numerous studies have mapped the WWW based on hyperlinks between websites (De Maeyer, 2013), and these consistently find that websites based in the US (and other G7 countries) receive links from websites based in the rest of the world, but few Western sites link to sites in the non-Western world (e.g., Park, Barnett & Chung, 2011). Likewise a large proportion of websites in most other languages link to sites in English, but few English language websites link to other language sites (Daniel & Josh, 2011). This literature considers a country's position on the hyperlinks graph a proxy for its potential for development, and predictors of their interactions with other countries (e.g, Schumate & Dewitt, 2008; Park et al., 2011). Studies find similar results from mapping the WWW using connections between countries based on Internet Service Providers (Ruiz & Barnett, 2014) and webpages' content (Chung, Barnett & Park, 2013). These maps, consistent with other representations of the global digital divide (e.g., Graham, 2014) suggest that the Internet replicates the modern world system, with the Western world at its "core" and the rest of the world at its "peripheries." Similarly, representations based on global Internet governance by transnational civil society organizations (e.g., Reporters without Borders) typically highlight the "balkanization" of this global medium due to nationally imposed regulations such as access blockage, and developing countries that are not formal liberal democracies are portrayed as the Internet's "black holes."[2]

All these imageries of the Internet - based either on technological infrastructure or content—instantiate a "media-centric" view of communication. When communication is premised on media technologies, there underlie unwarranted presumptions about usage. For instance, a hyperlink may exist but few users may click on it (Wu & Ackland, 2014)[3]. Providing more hyperlinks from Western sites to non-Western sites—making them less peripheral in the process—does not guarantee actual information exchange between the West and the Rest. Therefore, a core-periphery hyperlink structure does not equal to asymmetric information flows (see Barnett & Park, 2014). In a similar vein, certain governments may block some foreign websites, but most users may not access those sites even if they weren't blocked. What these media-centric maps share in common is not factoring, but taking technical features as proxies for, actual Internet use usage. They tend to imagine the WWW as a global medium upholding a constellation of undifferentiable masses and thus spaceless and without geography (Graham, 2013, 2014).

Media-centric representations also allow certain ideological bias, in particular, the championing of "Internet freedom" in US-led Western policy-making and popular media discourses. Importantly, Internet freedom here refers to access without political interference (Cramer, 2013; Mansell, 2014), and Internet governance debates thus center on state censorship, especially state imposed access blockage, practices associated with countries with political systems at odds with the US ideal of liberal democracy (McCarthy, 2011). Therefore in the US-led popular understanding, these countries are the problematic troublemakers at the margin, forming "black holes" that obstruct the realization of a web "without geography" (e.g., Reporters without Borders, 2007).

Some emerging scholarship has begun to recognize the limitations of media-centric imageries and discussions of the WWW. Without mapping actual user activities one cannot envision changes in the WWW resulting from the rapid online adoption of the vast populations' from the "peripheral" Global South. For example, Burrell (2012) contests the influential vision of the WWW as global system with a "universal digital language" (Castells, 1996, p. 2) and points out that the "Internet's interoperability rests on what is more aptly labeled a technical protocol," which is "not the same thing as a human language" (p. 185). During her ethnographic study of Ghanaian Internet use, Burrell (2012) realizes that "as divergent social groups go online, they make over the cultural topography of the Internet unevenly and heterogeneously" (p. 185). Barnett & Park (2014), finds the web to be an aggregation of many "small world" networks based on language, geography and history a trend which resembles how the world television landscape, once dominated by US produced imports, now appears as a mosaic of vibrant "culturally defined markets," rooted in regional production centers (Straubhaar, 2007). What is perceived as politically coerced balkanization of the Internet is in fact a fuller expression of linguistic and geographic boundaries, a consequence of the global spread of web use (Taneja & Wu, 2014).

Further, a media-centric perspective that presumes media use upon technical access is even more limited for the web than for traditional media. This is because, unlike in the days of broadcast media when people were left with limited and spatially-bounded options, today having access to the Internet means they could consume content from any corner of the WWW, irrespective of where they reside. Hence, having mere technological access to the "whole web" does not automatically imply that people would access everything they can; rather, users inhabit a rather narrow slice of the WWW shaped by various social structures. It is more inaccurate than ever to treat technologies as "proxies for the knowledge people draw upon in order to make sense of their world" (Mansell, 2013, p. 15) and to attribute to technical access the lack of use (e.g., of the Western web) at analytical levels such as countries or societies.

To summarize, when explaining global online communication, we propose a shift in focus from Internet infrastructure such as hyperlinks (i.e., purported proxy of global connectivity) and state-imposed blockage (i.e., denial of access to foreign information through firewalls) to people's actual web use globally—especially in non-Western societies. In particular, our empirical examination of web usage by non-Western communities avoids ascribing an a priori preference for their connection to the "world," which in the dominant discourse essentially implies the Western web. To "reimage" the Internet, we introduce methodological innovations to not only capture the online presence of various communities worldwide but also to gauge, separately, the relative location and level of horizontal connectivity within such communities.[4]

**Ethnological Mapping of the WWW: Distance and Thickness of Online Regional Cultures**

As the foundation of anthropology, ethnology is concerned with the characteristics of different cultures measured as "clusters of common concepts, emotions and practices" (de Munck & Korotayev, 2000, p. 338) and the relationships between them. Depending on the research questions, ethnologists study the corresponding "traits that may be distributed among the cultures that are included in the sample" (p. 347) and seek to evaluate and explain the "similarity or dissimilarity" between, as well as "the distinctiveness of" these cultures (Thomas, 1989, p. 27). Although nominally ethnology became marginalized in anthropology, ethnological thinking remains embedded in hidden assumptions of many research endeavors that aim to generalize and compare because it characterizes how human beings understand cultures, identities, and the world (MacRae, 2006, p. 118).

Our ethnological mapping of the WWW is based on actual web usage and it consists of three steps. A major obstacle to ethnological analysis is finding like entities to compare (MacRae, 2006), as cultural or social practices, discourses and even material objects, rarely come ready made for large-scale systematic comparison. Hence, we first identified traffic to the globally most used websites as a usage-based trait distributed among the plural cultures that reside within the global WWW. Accordingly, to gauge cross-cultural similarity (dissimilarity) we rely on shared traffic between websites measured by audience duplication, a metric we describe in the following section.

Having identified an appropriate measure of shared web traffic, our second step involved developing an analytic technique to determine the various online cultures as distinct “clusters of practices.” For ethnological cross-cultural comparisons, it is important to refrain from assigning units of culture based on a priori attributes other than the traits used to arrive at them (Munck & Korotayev, 2000). Our units of culture are essentially social structures we identify through a user-centric network analysis of shared traffic between websites, an approach that we adapt from Webster and Ksiazek (2012). Taneja and Wu (2014) refer to such structures as “culturally defined markets,” which are clusters of “culturally proximate” websites. We avoid the term “markets” due to the associated commercial overtones and instead interpret these clusters using the conceptual lever of “online regional cultures,” as the concept of “region” is theoretically and empirically consistent with the ethnological method. Conventionally, regional cultures are characterized by place-based language traditions and other cultural commonalities arising from shared climates, natural resources, and historical experiences (Griswold, 2008). Emerging empirical research on global online user behavior indicates that online cultures too have their origins in specific physical places (e.g., Taneja & Wu, 2014; Takhteyev, Gruzd & Wellman, 2012; Hecht & Gergle, 2010), which justifies our accent on the term “regional” when identifying cultures associated with the usage of the (potentially global) WWW.

Moreover, a region’s designated geography may not be congruent with a sovereign territory but instead above, below, or intersecting it ; it has fuzzy boundaries dependent upon the viewer's perspective (Griswold, 2008), which inevitably implicates the ethnological thinking of generalization and comparison (MacRae, 2006). For example, the region associated with the music created by singers from the holy Indian city of Varanasi could be Varanasi, the North Indian Plains, India, or even the Hindustani speaking South Asia comprised of parts of North India, Pakistan and Afghanistan. In addition to such contextual malleability, geographers increasingly conceptualize places "as constellations of connections that form, reform and disperse in space and over time" (Cosgrove, 2008, p. 47). All these characteristics lend the concept well to a network analytic approach, which simultaneously accounts for both the inherent and the relational traits of entities (Monge & Contractor, 2003)—a perfect iteration of ethnological thinking in examining the WWW. In addition, representing the web using network analysis avoids the construction of essentialized ethnological maps, a common pitfall in ethnology where boundaries are drawn based on static traits.

Having identified clusters of regional cultures, the final step is to measure their various characteristics in relation to each other and attempt to theoretically explain their emergence in the real-world. This step involves locating, at a specific point in time, online regional cultures in relation to one another by measuring their relative distances and discerning the strength or thickness of each. To begin, we map each culture's relative location using force-directed network visualizations; these place websites sharing more traffic closer to each other and enable us to "see" distinct cultures identified through formal cluster analysis.

Next, for cross-comparison, we measure how far a regional culture (cluster) stands from the rest of the world. We term this measure "distance" (a more neutral term in place of "isolation," used by Taneja and Wu (2014), which conventionally bears negative connotations of being backward and unsettling). Distance may result from political interventions such as access blockage. It may also be a consequence of the local society operating in a cultural system at a high “cultural distance” —a concept introduced by Hoskins and Mirus (1988) to calculate pair-wise distance between countries for studying international trade of cultural objects and other related aspects of globalization. Cultural distance reflects commonalities between cultures based on many dimensions such as language, religion, political system, and ethnic/racial composition (Shenkar, 2001), making the concept highly compatible with that of regional cultures, which are distinguished based on cultural commonalities. For example, the specific cultural and linguistic affinities established via former colonial ties may make the online regional culture rooted in India and Pakistan less distant from the rest of the WWW than one that originates in Japan.

Distance or the lack of it may not necessarily mean that the local communication landscape in point exhibits rich and meaningful connectivity within itself. Hypothetically, if China pushes access blockage to an extreme and makes its online regional cultures as distant as that of North Korea, China’s online regional culture, housed by the Internet infrastructure already ridiculed as the “Chinternet,” may nonetheless exhibit rich internal vibrancy and diversity (Taneja & Wu, 2014), whereas North Korea’s Intranet, also characterized by the lack of international linkages, provides meager online resources for its users (Warf, 2015).

To assess this separate trait of an online regional culture, we introduce the concept of thickness. Initially used in critical transcultural media research, the concept of “thickening” refers to the intensification of patterned connectivities “at the level of thinking, discourse, and practices” that distinguishes media cultures from one another (Couldry & Hepp, 2012, p. 99). Pertinent to our approach, a media culture’s thickness is judged by people’s habitual usage of various media offerings associated with it, in addition to their content and formats (p. 101). It is individuals’ patterned usage in fulfilling their economic, cultural, political, and leisure needs that fundamentally structures and consolidates the local communication landscape (Couldry & Hepp, 2012). In this light, by thickness of an online regional culture, we mean the level of concentrated reliance on the set of web content that represents this regional culture. The growth of locally relevant cultural products and e-services, the expansion of a region's online population which usually accompanies a bridging of the digital divide and the prosperity of online interaction between users associated with this regional culture, all potentially contribute to thickening. Examining such links, our ethnology of the WWW provides insights on the empirical consequences of various factors on patterns of web usage globally.

## Method

### Creating a Cross-cultural Dataset

We analyzed shared user traffic between the world's top 1000 web domains at three time points (September of 2009, 2011 and 2013). We obtained web usage (traffic) data from comScore, a panel-based service that provides metered Internet audience measurement data once a month from 2 million users worldwide in 170 countries. We obtained from comScore traffic data for the top 1000 web domains globally (ranked by monthly unique users), as these domains not only attract 99% of all the web users at all times but also ensure an adequate representation of different languages and different geographies. ComScore reports data on websites at the level of web domains as well as for subdomains for many large websites. This in part depends on the architecture of the website. Google, for instance, has its different geo-linguistic variants as separate web domains (e.g., www.google.es, www.google.de, etc.) and comScore reports them as such. On the contrary Wikipedia language versions are sub-domains of the main domain (e.g., es.wikipedia.org). In such cases, we considered these sub-domains as separate websites in our final sample instead of including only higher level domains. We captured these data at three time points, September 2009 (1018 websites included in the final sample), September 2011 (1022 websites included), and September 2013 (1031 websites included).

As already noted, we draw on web traffic as a shared trait of online users across cultures and identifying shared web traffic allows cross-cultural comparisons. To facilitate such analysis, we utilize a measure called “audience duplication,” which simply put is the extent to which two media outlets (e.g., websites) are consumed by the same set of people in a given time period. In a hypothetical universe of 100 people, if on a given day 20 people accessed both CNN.com and Baidu.com, the audience duplication between these two websites would be 20 or 20%. For all possible pairs of media outlets under consideration, calculating duplication in this manner results in a symmetric audience duplication matrix, which can then be partitioned to identify communities of media outlets with similar audience duplication patterns.

Generally, any two websites have some audience duplication, which may be due to random chance. Therefore we calculated a measure called *above random duplication (*also used by Webster and Ksiazek, 2012), which is the audience duplication between two websites after accounting for expected duplication due to chance. As an example, if in a given month 80% of all web users visit Facebook.com and 70% visited Google.com, the expected duplication between them would be 56% (80% * 70%). The duplication of interest then is a measure termed “above random duplication,” the residual value obtained by subtracting the observed duplication from the expected duplication. The resulting matrix with above random duplication as the cell value can be analyzed as a *valued network* (negative values are treated as zero; in other words, ties are considered absent). These values can also be converted to *dichotomized data* by regarding any positive value of above random duplication as ‘1’ (tie present) and a zero or negative value as ‘0’ (tie absent).

For each website in our sample, we obtained its audience duplication with all other websites from the same annual sample. Thus in the final dataset, we have 517,653 ((1018 *1017)/2) pairs of audience duplication in 2009, 521,731 ((1022*1021)/2) pairs in 2011, and 529,935 ((1030*1029)/2) pairs in 2013. Using the method described above we calculated the above random duplication for each pair and then obtained two sets of matrices (graphs), one with valued ties and the other with dichotomized ties. A coder visited each website to note all the languages in which the website offered content.

## Analysis

### Force-directed Network Visualization: Ethnological Mapping of WWW usage

For ethnological mapping, we employed the Fruchterman and Reingold (1991) visualizing algorithm, which belongs to a class of visualization techniques known in graph theory as force-directed graphs. The basic mechanism is that there are repulsive forces between all nodes and nodes with edges have attractive forces and hence are placed topologically adjacent to each other. Therefore, in the final visualization, all nodes that tend to have ties with one another are placed in close vicinity to form tightly knit clusters; such clusters are relatively separated from other similarly formed clusters. Figure 1(a), 1(b) and 1(c) represent the visualizations of the 2009, 2011 and 2013 graphs, where the dots are the nodes (websites) and the lines represent (dichotomized) ties between them based on audience duplication. These “show” fair evidence that global web usage clusters itself into many communities of websites based on similarities in audience duplication. We conducted cluster analysis to identify these communities, which confirmed them to be expressions of online regional cultures.

**FIGURE 1 ABOUT HERE**

### Cluster Analysis: Identifying and Measuring Online Regional Cultures

The clustering coefficient of a graph indicates the presence of tightly knit groups of nodes that have disproportionately high ties with each other relative to nodes in other groups.

Theoretically, this varies between 0 and 1 and the clustering coefficient of a random network is equal to its density and a value higher (lower) than the density indicates more (less) clustering than a random graph of the same size and density. Our datasets had a density of 0.43, 0.40 and 0.35 and clustering coefficients of 0.86, 0.85 and 0.86 in 2009, 2011 and 2013 respectively. Therefore, overall the network had a high tendency to cluster into subgroups, which increased over time (the clustering coefficient remained high even as the density reduced). Thus websites clustered into groups based on high audience duplication among sites within the group and relatively low duplication between sites that belong to different groups.

We divided these graphs (with valued ties) into clusters using a hierarchical clustering approach, using Pearson correlation coefficient between any two pairs of sites as measure of similarity. In other words, a high correlation between any two sites suggests that they tend to share audiences with the same sets of sites. Consequently, groups of sites with high inter-correlations tend to cluster together. Simply put, these clusters are groups of websites that share audiences with one another to a greater extent than with websites outside their groups. Further, we assigned each cluster with a unique color for easy identification, consistent across all visualizations.

Following the same approach used by Taneja and Wu (2014), we estimated the *distance* of each cluster from the rest of the WWW using a basic node-level network measure termed *closeness (or farness) centrality*, an indicator of the average closeness (or distance) of a node from all other nodes in the network. Each cluster in our WWW networks is basically a group of nodes. In order to calculate the distance of each cluster, we treated the entire cluster as one node, and computed its average shortest distance from all other nodes in the network. We calculated this distance for each cluster we observed in 2009, 2011 and 2013. A high *distance* score suggests that even users of a cluster's most well-connected (i.e., widely used) website tend to engage with fewer websites outside the said cluster compared to a similar website in a cluster with a lower distance score.

Distance as measured here indicates little about the connectivity within a cluster, or its thickness. For estimating the *thickness* of each cluster, we utilize a measure called the E-I index, which is typically used in social network analysis to indicate relative cohesion within social groups (Krackhardt & Stern, 1998). It is calculated as the ratio of the difference and the sum of external and internal ties of a cluster. Thus a cluster with only one actor can only have external ties and hence its E-I index would be ‘1,’ indicating no cohesion. On the contrary, a cluster of nodes with all internal ties and no external ties would have an E-I score of ‘-1,’ indicating perfect cohesion. If these usage-based clusters indeed represent online regional cultures as we expect, we believe that an E-I index can be used to observe the intensity of the intra-community connectivity, or the “thickness” of the local regional culture. We use the valued ties to obtain these scores to account for the volume of shared audience traffic between sites rather than just the presence or absence of a connection.

Although E-I scores indicate relative cohesion of clusters within a given network, in our present case, changes in relevant E-I scores provide evidence of cultural thickening over time. This is because the graphs generated for the three different time points share substantive similarities and, across the years, we are able to identify a homologous set of major clusters—that is, clusters representing the same regional cultures over time. Since most clusters in our data have all possible internal ties (due to the design of the clustering algorithm), a reduction in the E- I index indicates relative decrease in connections between nodes within and those outside the cluster. In terms of global web use, this suggests that from 2009 to 2013, the extent of audience duplication among sites within a regional culture (cluster) increased relative to their duplication with sites in other cultures (clusters). This indicates a declining tendency of the average regional sites’ user to visit websites unaffiliated to the said region culture. If accompanied by an overall increase in the number of sites affiliated to this regional culture, a reduction in E-I index indicates cultural thickening.

**FIGURE 2 ABOUT HERE**

In Figure 2, we report these key measures associated with each of the major clusters. Each graph is a scatterplot with the distance score in the X-axis and the thickness (E-I index) score in the Y-axis. The diameter of the circle corresponds to the cluster size, measured as the number of websites in that cluster. Figure 3 shows the relative E-I scores (standardized) across a few clusters that were comparable across the year 2009, 2011 and 2013.

**FIGURE 3 ABOUT HERE**

## Contextualizing the Evolution of Online Regional Cultures

In this section, we interpret the evolution in global user behavior in relation to local cultural identity, state intervention, and larger economic and social processes.

### Web User Behavior and Cultural Identity

In all three years, we find that the WWW as a network of shared usage between websites largely clusters on linguistic and geographical lines, affirming the importance of local cultural identity stressed in previous literature. To begin, we notice different categories of clusters dependent on the interaction between language and geography. This is consistent in each of the three years' traffic data we analyzed and in line with findings from other recent studies using similar traffic data collected at a different time point (Barnett & Park, 2014, Taneja & Wu, 2014). State imposed access blockage plays at best a limited role in the formation of these fault lines.[5] In many cases, clusters could be linked explicitly to contiguous geographical spaces defined by nation states such as China (which sometimes coalesce with sites on Taiwan and Hong Kong),[6] Poland, Korea, Italy and Japan where the principal language is exclusive to the geography. We also witness distinct clusters for certain countries (such as United Kingdom and Canada) despite their languages being spoken very widely in many other countries. In other cases, websites focusing on multiple countries cluster together based on a common language, as seen for Spanish language sites from Spain, Mexico and Argentina. Yet another category of geo-linguistic clusters we observe is comprised of global sites (such as Facebook, Youtube and Ikea) that are present in multiple languages and have specific variants focusing on different countries. Finally, we find some evidence of genre based clusters, with two salient examples being porn sites and file/video sharing platforms. Such websites rely minimally on language and hence apparently transcend regions more easily.

Our analyses further reveal the primary role of language in the landscape of horizontal communication online. First, the more specific a cluster's linkage to a language, the more distant it is from the rest of the WWW. We find that the distance increases if a language is exclusive to geography. For instance, clusters corresponding to Japan and Korea are consistently more distant than the ones corresponding to Spanish or English speaking sites. The cluster which we term as "US / Global" in 2009 is the least distant from the rest of the WWW, due to the presence of many "global" sites in this group. However, the distance of the cluster of US-focused sites increases in 2011 and 2013 (where we accordingly term this cluster "US / English") as in these years, many of the "global" sites (such as Facebook and Twitter), probably as the proportion of non-US user participation increases, separate from the US cluster to form a cluster of their own. There is an overall increase in cluster distance in 2013, which is consistent with the finding that compared to previous years, the network is more clustered in 2013 (the least distant cluster in 2013 has a higher distance score than that of its equivalent in both 2009 and 2011).

Relatedly, a regional culture formed with one primary language is likely to be more distinctive, as well as thickens faster, than one formed in a geographically contiguous region with linguistic plurality. For instance, in our analysis, certain clusters such as that of Spanish language sites and that rooted in former USSR countries have formed mainly due to shared regional languages transcending geographic or political boundaries. We also observe the opposite for India where linguistic plurality, despite geographic contiguity, has probably impeded the process of thickening among an otherwise large online regional culture. That the region has not thickened as much as others with a comparable user base may be partly explained by the fact that, although being the primary (and common) language of most sites in the Indian cluster, English is only spoken by upper class Indians that make up about 10% of the total population.

Another important observation is that, although major geo-linguistic clusters are largely consistent across years, our analyses show an overall increase in the number of distinct regionally oriented online cultures, resulting from the relative expansion of the web usage in the global South. This trend holds for China, India, Brazil, and former USSR countries: the Chinese cluster, which nearly doubled in size between 2009 and 2013, showed the most phenomenal growth. Given the similar sample size in each year (approx. top 1000 sites), the growth of certain clusters is balanced by the decline in others. The Korean cluster, while prominent in 2009 and 2011, is missing from the 2013 visualization (it had 34 sites in 2009 to 33 in 2011 and only 4 in 2013). Germany, France and Japan are other prominent clusters that shrank in size. In general, we notice that regions where Internet penetration has grown significantly during the observation period witness cultural thickening online. This is a likely outcome of local users' increasing reliance on regional websites, which manifests as intensification of horizontal communication within the regions, when seen in the aggregate. Likewise, the cluster of US-based sites also thickened once they segregated from the major "global" sites into a cluster of their own. The stable or increasing thickness scores of clusters corresponding to regions with stagnant Internet population numbers such as Korea, Japan, Germany and France support this interpretation, as the proportion of web users from these countries most likely declined from 2009 to 2013 in our sample.

Connected to the Internet penetration rate is the social and economic status of the regional online population and its level of regional orientations. In developing countries, Internet use is a privilege of the elite segment marked by higher cultural capital for navigating and making sense of cultures outside of the region. In these countries, as the Internet penetrates down the social ladder, it accommodates more users who likely lack the capacities of and/or interests in making use of foreign websites. Thus over time the growth of the percentage of users from lower strata is associated with the emergence of distinct regional orientations, which simultaneously contributes to the distance and thickening of the online regional cultures in point. In 2013, for instance, India and Japan both had about 100 million Internet users each. Yet, our data show Japan has a far thicker online culture than India. This is because Indian users who constituted the top 10% of the national populace belonged to very privileged social groups that are at ease with the established “global” and English cyberspace. As Internet usage picks up among India’s lower social strata, newer regional cultures based on India’s many languages may emerge over and above the currently visible national culture relying on the English lingua franca.

**Web Use Behavior and Local Political Economy**

A closer look at the thickening trajectories of various online regional cultures in relation to state policies and other economic and social contexts suggests the critical relevance of local political economy to web use behavior. Such relevance can be understood in terms of, on the one hand, the structural environment facilitating local *consumption*—that is, to enable more members of the community to engage with cyberspace. The uneven Internet penetration levels are themselves affected by state policy. For instance, in South Korea government policy has been instrumental in achieving near ubiquitous broadband access that is both fast and of low cost. In stark contrast to South Korea, Internet user base in Brazil and India has taken much longer to grow. Brazil had only 24 million Internet users (13% of its population) in 2003 (compared to Korea at more than 60% in the same year), and it surpassed 100 million Internet users (50% of its population) only in 2013, a fact consistent with the growth of the Brazilian cluster in 2013 seen in our data. Much of the growth in the Brazilian Internet user base in the last 3 years seems to be from Internet use on mobile devices. In 2004, India had less than 2% of its population online. In 2013, even though only 10% Indians are online, this corresponds to about 120 million users.[7]

On the other hand, the politico-economic contexts also affect the dimension of local *production*, or inclusion of local communities in designing and producing content that populates the part of WWW they inhabit. In this regard, we notice the connection between the changes in thickness scores and the conditions of homegrown Internet industries. For local horizontal connectivity to develop, Internet users should grow in tandem with online content focusing on the region and web tools that enable people to discover such content. Generally speaking, local producers are more successful in fulfilling such needs, especially when embedded in a healthy innovation culture that is supported by friendly state policies. Taking South Korea again as an example; its high Internet penetration is supplemented by an innovation culture fostering startup enterprises. Launched five years before Facebook, CYworld remained South Korea's most popular SNS for a decade (The Economist Explains, 2014). The cultivation of China's vivacious online regional culture can also be attributed to its robust local Internet industry and the key presence of various government interventions. Chinese users have witnessed the development of corporations such as Tencent technologies which, now leading competition worldwide in innovations on the mobile web, far preceded US-based corporations in creating innovative digital platforms for online social networking **(**Elliot, 2014**)**. Likewise, Japan, Russia and Brazil all have massive local Internet portals that cater to regional needs.

The preceding discussion on the relations between the web use of local societies and the political economies governing these societies also provides a timely critique of the dominant “Internet freedom” rhetoric, which in effect adds to the entitlement of multinational Internet companies that demand free operation both in the US and abroad without government intervention (Cramer, 2013). In fact, media policy scholars suggest that US foreign policies serve the interests of large corporations by equating the goal of “liberalizing foreign polities with the values of human rights and democracy” to that of “opening [foreign] markets to US capital,” (McCarthy, 2011, p. 89; also see Nordenstreng, 2011; Schiller & Sandvig, 2011). This one-dimensional conception of Internet freedom as the absence of political intervention explains why US policy-making and popular media discourses have little recognition of the role of active government interventions, particularly in numerous developing states, in providing Internet access and local content. Similarly, while obsessed with the detail and scales of state imposed blockage (Taneja & Wu, 2014), they are indifferent to Internet companies’ targeted blockage of users from poor countries as the latter may adversely impact their revenues (Burrell, 2012).

In contrast to the prevalent emphasis on political constraints, our study draws attention to the generally ignored but significant economic constraints to the global inclusion in online communication (see also Cramer, 2013). Portraying the online activities and cultures of developing countries in a more balanced light, it foregrounds the development and ecology of online regional cultures based in the global South and recognizes that in many cases it is local government interventions that eliminate economic barriers, which in effect expand the Internet freedom of their citizenry. For instance, certain initiatives and regulations may foster high growth in both domestic usage and homegrown Internet industries, thus enabling more people in these countries to benefit from rich lives on the Internet and to at once shape their own online habitat. People thus get to enjoy indigenous cultural content, interact with their communities, and utilize various e-services tailored to the locality. This is consistent with a UNESCO-directed study of multi-country data, which suggests strong correlations between the development of local network infrastructure, the lowering of access prices, and the growth of local content (Internet Governance Forum, 2014).

## Conclusion

Our study contributes, conceptually and methodologically, to the research, documentation and evaluation of the development of the WWW. Through a network analysis of shared global traffic between 1000 most popular websites at three time points, we have advanced the approach of ethnological mapping to better understand worldwide web use and its association with local cultural identity and politico-economic contexts. With methodological innovations tailored to data of a new nature, our approach generates a new imagery of the WWW that foregrounds usage and its shaping forces, in contrast to the existing imageries that are based on technical features of the Internet. It also provides granular measures for exploring local communities' engagements with the Internet.

Our major findings are as follows. First, the locations of online regional cultures in relation to one another and with their underlying geo-linguistic linkages lend support to the common perception of cultural adjacency. Second, regional orientations--that is, tendency to use webpages in local languages--strengthen as Internet use expands beyond globalized elites. Third, in terms of intra-community vis-à-vis inter-community connectivities, online regional cultures associated with the global South have significantly grown and thickened.

To the best of our knowledge, our analysis is the first imaging exercise of the WWW on a website level at a global scale that conceives and demonstrates that "the Internet is not an abstract space or digital global village, but rather a network that enables selective connections between people and information" (Graham, 2013, p. 180). Geographers emphasize that, "representations of place are never neutral or objective and are always created in order to serve particular purposes" (Graham, 2014, p. 103). Compared to the prevalent media-centric Internet maps, our user-centric ethnological mapping method stands to challenge, rather than reinforce, the status quo presumption of a WWW that is anchored by Western knowledge, norms, and activities. It encourages the (Anglophone, especially) general public to confront the narrowness of the online world with which it is familiar. Further, the trend our maps capture may stimulate popular imaginations to envision a future cyberspace of vast heterogeneity, where one needs to get out of one's way to comprehend.

Moreover, these user-centric maps may inform policy-makers around the world in deliberating how to better empower the global South by introducing political, economic, and social opportunities with new technologies. By linking our empirical findings about online regional cultures to contextual indicators, we draw attention to the role of cultural identity, affirmative state intervention and the economic conditions of various countries in the formation of and changes in the global web use landscape. Our ethnological maps provide powerful visuals that support previous analyses in this realm; in recent years in international Internet governance discussion, this corpus of evidence has been the basis for the growing advocacy for affirmative government intervention in fostering local production of web content (e.g., Internet Governance Forum, 2014). In this sense, our analyses not only provide substance to this advocacy, but also serve as an initial step to move beyond concerns for cultural, particularly linguistic, diversity in web content to also address diversity in the dimension of web usage (see Napoli & Karppinen, 2013). Finally, our imaging exercise gives sustenance to a conception of Internet freedom broader than what is commonly conceived in US-led Internet governance discourse. This broader conception emphasizes the elimination of not only political, but also economic barriers for users around the globe so that they may shape and make use of online content they find appealing, contributing to the horizontal connectivity of our expanding WWW.

---

[1] We use "infrastructure" to refer to a range of levers of Internet architecture such as ISPs, routers, protocols, under-sea cables, hyperlinks etc, that enable or constrain usage. This is consistent with the use of the term in the literature (e.g., DeNardis, 2012; Dodge, 2008)

[2] Another type of popular WWW imageries is based on the coverage of usually US-based IT giants such as Facebook and Twitter, which veils the political and economic specificities of the particular online applications (see Menchen-Trevino, 2013).

[3] Additionally, hyperlinks may contain erroneous links, irrelevant information and inconsequential relationships (Weber & Monge, 2011)

[4] Taneja and Wu (2014) direct scholarly and public attention from its conventional focus on China's governmental filtering mechanism to the web usage under its local influences. This study suggests no clear role of access blockage in explaining the extent of isolation of Chinese web browsing behavior.

[5] To be clear, we do believe that the fast inclusion of users from and of websites focusing on formerly underrepresented regions online may cause changes that conflate with the effects of governmental blockage

[6] It is hard to tell from our data whether the Chinese cluster has users only from mainland China or from HK and Taiwan as well. However, we see a separate cluster of Taiwanese sites including the Chinese Wikipedia in 2013.

[7] All Internet penetration figures are from World Development Indicators from World Bank, made available by Google Public Data Explorer.

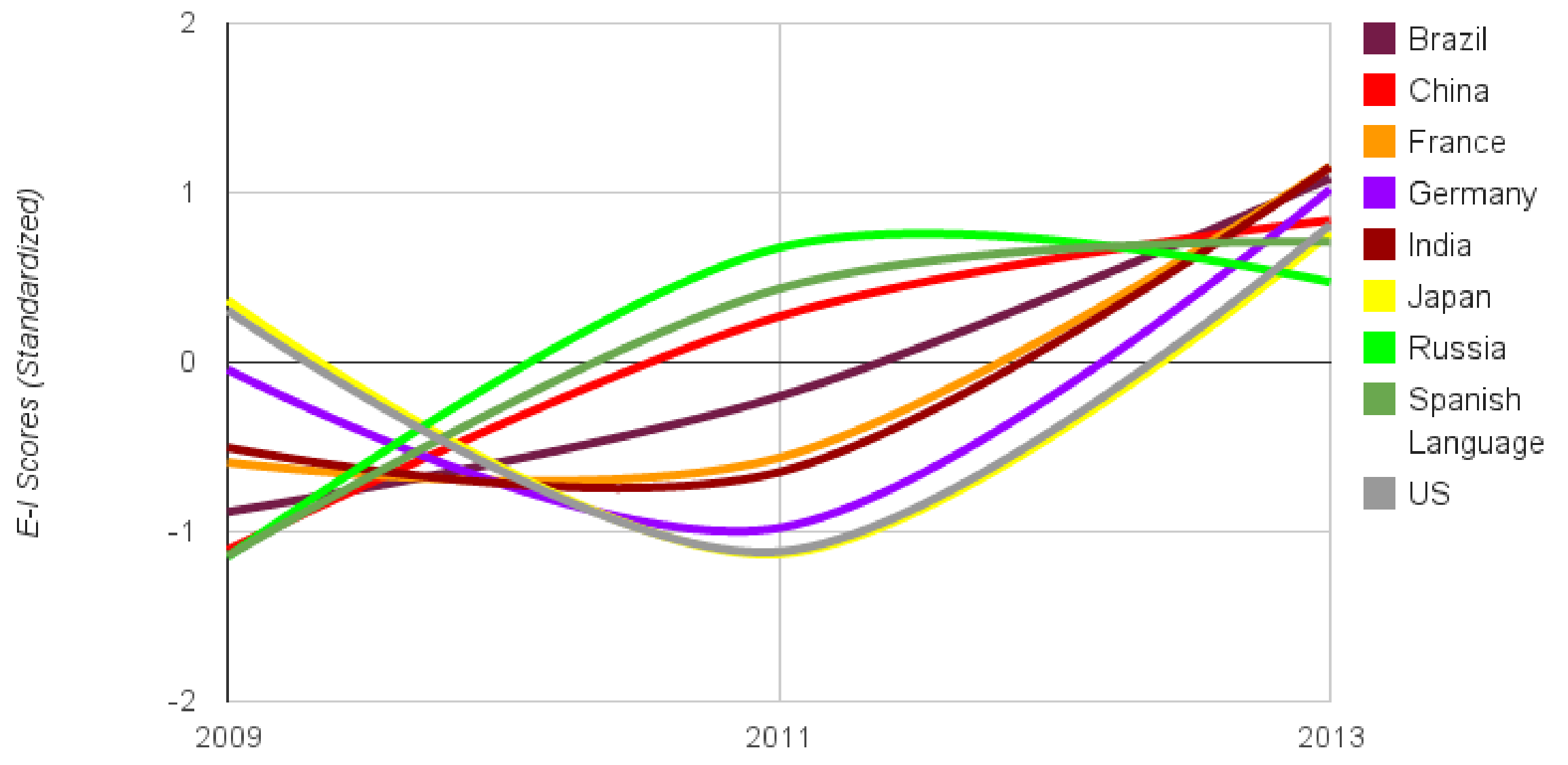

Figure 3 Thickening of Online Regional Cultures over Time

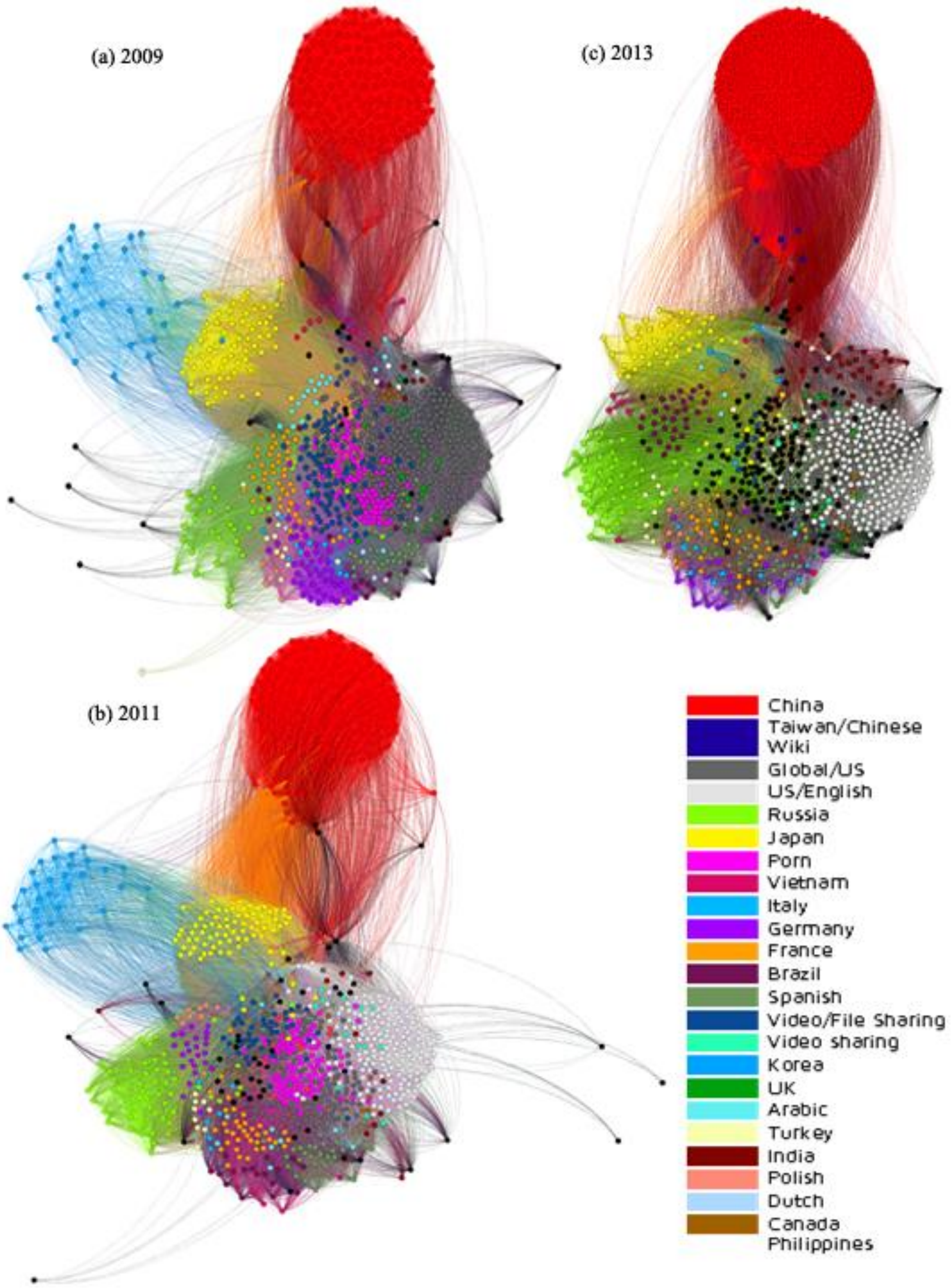

Figure 1: Ethnological Mapping of Global Web Use

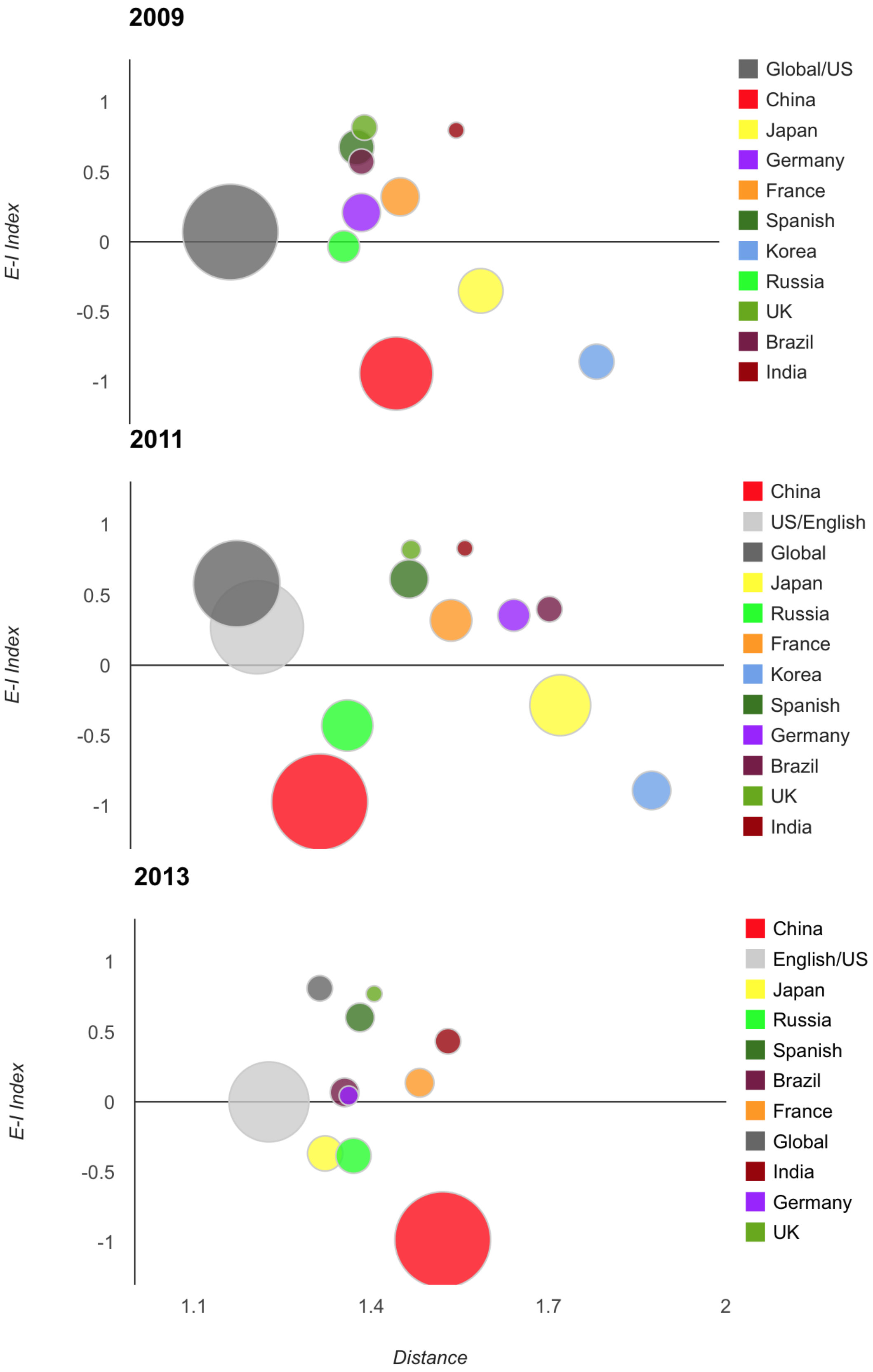

Figure 2 Evolution of the WWW (Distance and Thickening)

**About the Authors**

**Angela Xiao Wu** (PhD, Northwestern University) is an Assistant Professor in the School of Journalism & Communication at the Chinese University of Hong Kong. She explores the connection between emergent technologies and cultural changes using mixed methodologies. **Address:** Room 206-7, Humanities Building, Chinese University of Hong Kong, Shatin, N. T., Hong Kong. **Email:** wu.angela.xiao@gmail.com

**Harsh Taneja** (PhD, Northwestern University) is an Assistant Professor in the School of Journalism at University of Missouri. His research focuses on media use, specifically how contextual and individual factors shape patterns of audience formation. **Address:** 181-C Gannett Hall, Missouri School of Journalism, Columbia, MO, USA -65211-1200. **Email:** harsh.taneja@gmail.com